\documentclass[aps,pra,superscriptaddress,floatfix,showpacs,10pt, twocolumn]{revtex4}
\usepackage[dvips]{graphicx}
\usepackage{amsmath,times}

\begin{document}

\title{Generation of cluster states with Josephson charge qubits }
\author{Xiao-Hu Zheng\footnote{Electronic address:
xhzheng@ahu.edu.cn}, Ping Dong, Zheng-Yuan Xue and Zhuo-Liang
Cao\footnote{Electronic address: zlcao@ahu.edu.cn (Corresponding
Author)}}

\affiliation{Key Laboratory of Opto-electronic Information
Acquisition and Manipulation, Ministry of Education, School of
Physics and Material Science, Anhui University, Hefei, 230039, P R
China}

\begin{abstract}
A scheme for the generation of the cluster states based on the
Josephson charge qubit is proposed. The two-qubit generating case is
first introduced, and then generalized to multi-qubit case. The
scheme is simple and easily manipulated, because any two charge
qubits can be selectively and effectively coupled by a common
inductance. More manipulations can be realized before decoherence
sets in. All the devices in the scheme are well within the current
technology.
\end{abstract}

\maketitle

Quantum entanglement plays one of the most important roles in the
quantum information processing. Many ingenious applications of
entanglement have been proposed, such as quantum dense coding
\cite{dense}, quantum teleportation \cite{teleport}, quantum
cryptography \cite{ekert}, \emph{etc}. In achieving the task of
quantum communication, entangled states are used as a medium for
transferring quantum information. Meanwhile, entangled states are
also used for speeding up quantum computation. Therefore, the
preparation of entangled states becomes a key step towards quantum
computation. Though bipartite entanglement is well understood, the
extensive researches of multipartite case are still difficultly
proceeding. For a tripartite-entangled quantum system, there are two
irreducible classes of entangled states \cite{trientangle}.
Recently, Briegel and Raussendorf \cite{Briegel} introduced a new
class of \emph{N}-qubit entangled states, \emph{i.e.}, the cluster
states, which have some special properties. In addition to the
properties of both Greenberger-Horne-Zeilinger (GHZ) and W-class
entangled states, they especially hold a large persistence of
entanglement, that is, they (in the case of $N>4$) are harder to be
destroyed by local operations than GHZ-class states. It has been
shown that a new Bell inequality is maximally violated by the
four-qubit cluster states, and isn't violated by the four-qubit GHZ
states \cite{Scarani}. More significantly, the cluster states are
regarded as a resource of multiqubit entangled states, thus cluster
states become an important resource in the physics, especially in
quantum information.

Recently, much attention has been attracted to the quantum computer,
which works on the fundamental quantum mechanical principle. The
quantum computers can solve some problems exponentially faster than
the classical computers. For realizing quantum computing, some
physical systems, such as nuclear magnetic resonance \cite{18},
trapped irons \cite{19}, cavity quantum electrodynamics (QED)
\cite{20}, and optical systems \cite{Turchette} have been proposed.
These systems have the advantage of high quantum coherence, but
can't be integrated easily to form large-scale circuits. As is well
known, the cluster states are mainly applied to quantum computing.
In Ref. \cite{Raussendorf}, Raussendorf and Briegel described the
so-called one-way quantum computer, in which information is written
onto the cluster and read out from the cluster by one-qubit
measurements. A number of applications using cluster states in
quantum computation have been proposed \cite{Zhou}. Thus the
preparation of the cluster states has become the focus of research.
Recently, Zou et al. proposed probabilistic schemes for generating
the cluster states of four distant trapped atoms in leaky cavities
\cite{Zou} and linear optics systems \cite{15}. Barrett and Kok
proposed a protocol for the generation of the cluster states using
spatially separated matter qubits and single-photon interference
effects \cite{16}. Yang et al. proposed an efficient scheme for the
generation of the cluster states with trapped ions \cite{17}. We
also proposed two schemes for the generation of the cluster states
via both cavity QED techniques and atomic ensembles \cite{Dong}.

As a solid-state qubit, Josephson charge qubit is one of  the
promising candidate for quantum computing. Accordingly, generation
of the cluster states by Josephson charge qubit is of great
importance. Josephson charge \cite{21-1,21-2, 22} and phase
\cite{23Mooij, 24} qubits, based on the macroscopic quantum effects
in low-capacitance Josephson junction circuits
\cite{Makhlin,You2005}, have recently been used in quantum
information processing because of large-scale integration and
relatively high quantum coherence. Some striking experimental
observations \cite{Nakamura,van} demonstrate that the Josephson
charge and phase qubits are promising candidates of solid-state
qubits in quantum information processing. In particular, recent
experimental realizations of a single charge qubit demonstrate that
it is hopeful to construct quantum computers by means of Josephson
charge qubits \cite{Nakamura2}. In this paper, we propose a scheme
for the generation of the cluster states using Josephson charge
qubit. This scheme is simple and easily manipulated, because any two
charge qubits can be selectively and effectively coupled by a common
inductance. More manipulations can be realized before decoherence
sets in. All of the devices in the scheme are well within the
current technology. It is the efficient scheme for the generation of
the cluster states based on the Josephson charge qubit.

\begin{figure}[tbp]
\includegraphics[scale=0.95,angle=0]{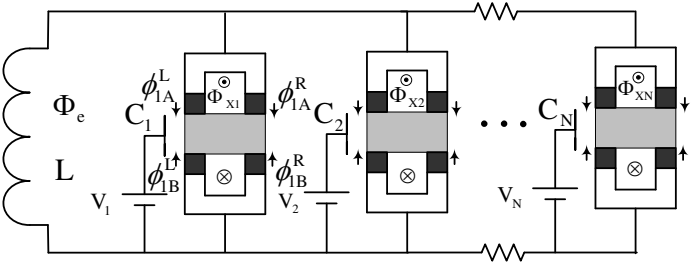}\caption{ Josephson charge-qubit structure. Each CBP is
configure both in the charging regime $E_{ck}\gg E^0_{Jk}$ and at
low temperatures $k_BT\ll E_{ck}$. Furthermore, the superconducting
gap $\Delta$ is larger than $E_{ck}$ so that quasiparticle tunneling
is suppressed in the system. } \label{fig1}
\end{figure}

Since the earliest Josephson charge qubit scheme \cite{21-1} was
proposed, a series of  improved schemes \cite{21-2,You2002} have
been explored.  Here, we concern the  architecture of Josephson
charge qubit in Ref. \cite{You2002}, which is the first efficient
scalable quantum computing (QC) architecture. The Josephson charge
qubits structure  is shown in Fig.(\ref{fig1}). It consists of
\emph{N} cooper-pair boxes (CPBs) coupled by a common
superconducting inductance L. For the \emph{k}th  cooper-pair boxe,
a superconducting island with charge $Q_{k}=2en_{k}$ is weakly
coupled by two symmetric direct current superconducting quantum
interference devices (dc SQUIDs) biased by an applied voltage
through a gate capacitance $C_{k}$. Assume that the two symmetric dc
SQUIDs are identical and all Josephson junctions in them have
Josephson coupling energy $E_{Jk}^{0}$ and capacitance $C_{Jk}$. The
self-inductance effects of each SQUID loop is usually neglected
because of the very small size ($~1\mu m$) of the loop. Each SQUID
pierced by a magnetic flux $\Phi_{Xk}$ provides an effective
coupling energy $-E_{Jk}(\Phi_{Xk})\cos\phi_{kA(B)}$, with
$E_{Jk}(\Phi_{Xk})=2E_{Jk}^{0}\cos(\pi\Phi_{Xk}/\Phi_0)$, and the
flux quantum $\Phi_0=h/2e$. The effective phase drop $\phi_{kA(B)}$,
with subscript $A(B)$ labeling the SQUID above (below) the island,
equals the average value $[\phi_{kA(B)}^L+\phi_{kA(B)}^R]/2$, of the
phase drops across the two Josephson junctions in the dc SQUID, with
superscript $L(R)$ denoting the left (right) Josephson junction.

For any given cooper-pair box, say $i$, when
$\Phi_{Xk}=\frac{1}{2}\Phi_0$ and $V_{Xk}=(2n_k+1)e/c_k$ for all
boxes except $k=i$, the inductance $L$ connects only the $i$th
cooper-pair box to form a superconducting loop, as shown in Fig.
2($a$). In the spin-$\frac{1}{2}$ representation, based on charge
states $|0\rangle=|n_i\rangle$ and $|1\rangle=|n_{i+1}\rangle$, the
reduced Hamiltonian of the system becomes \cite{You2002}
\begin{equation}
\label{1}
H=\varepsilon_{i}(V_{Xi})\sigma_z^{(i)}-\overline{E}_{Ji}(\Phi_{Xi},
\Phi_e, L)\sigma_x^{(i)},
\end{equation}
where $\varepsilon_{i}(V_{Xi})$ is controlled by the gate voltage
$V_{Xi}$, while the intrabit coupling $\overline{E}_{Ji}(\Phi_{Xi},
\Phi_e, L)$ depends on inductance$L$, the applied external flux
$\Phi_e$ through the common inductance and the local flux
$\Phi_{Xi}$  through the two SQUID loops of the $\emph{i}$th
cooper-pair box. By controlling $\Phi_{Xk}$ and $V_{Xk}$, the
operations of Pauli matrice $\sigma_z^{(i)}$ and $\sigma_x^{(i)}$
are achieved. Thus, any single-qubit operations are realized by
utilizing the Eq. (\ref{1}).
\begin{figure}
\includegraphics[scale=0.95,angle=0]{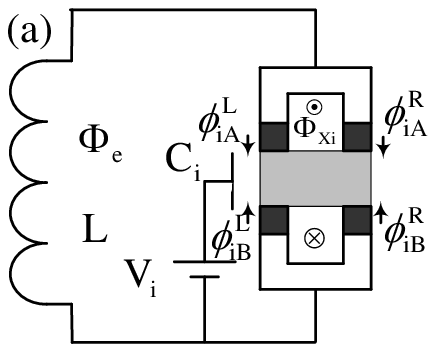}
\includegraphics[scale=0.95,angle=0]{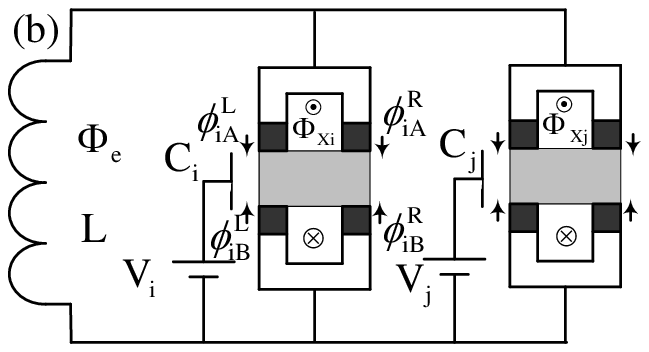}
\caption{(a) single-qubit structure where a CPB is only
 connected to the inductance. (b) Two-qubit structure where two CPBs are connected to the common inductance.} \label{fig2}
\end{figure}

To manipulate many-qubit, say $i$ and $j$, we configure
$\Phi_{Xk}=\frac{1}{2}\Phi_0$ and $V_{Xk}=(2n_k+1)e/c_k$ for all
boxes except $k=i$ and $j$. In the case, the inductance $L$ is only
shared by the cooper-pair boxes $i$ and $j$ to form superconducting
loops, as shown in Fig. 2($b$), the Hamiltonian of the system can be
reduced to \cite{You2002, You2001}
\begin{equation}
\label{2}
H=\sum_{k=i,j}[\varepsilon_{k}(V_{Xk})\sigma_z^{(k)}-\overline{E}_{Jk}\sigma_x^{(k)}]+\Pi_{ij}\sigma_x^{(i)}\sigma_x^{(j)},
\end{equation}
where the interbit coupling $\Pi_{ij}$ depends on both the external
flux $\Phi_e$ through the inductance $L$, the local fluxes
$\Phi_{Xi}$ and $\Phi_{Xj}$ through the SQUID loops. In Eq.
(\ref{2}), if we choose $V_{Xk}=(2n_k+1)e/c_k$, the Hamiltonian of
system can be reduced to
\begin{equation}
\label{3}
H=-\overline{E}_{Ji}\sigma_x^{(i)}-\overline{E}_{Jj}\sigma_x^{(j)}+\Pi_{ij}\sigma_x^{(i)}\sigma_x^{(j)}.
\end{equation}
For the simplicity of calculation, we assume
$\overline{E}_{Ji}=\overline{E}_{Jj}=\Pi_{ij}=\frac{-\pi\hbar}{4\tau}$($\tau$
is a given period of time), which can be obtained by suitably
choosing parameters. Thus Eq.(\ref{3}) becomes
\begin{equation}
\label{4}
H=\frac{-\pi\hbar}{4\tau}(-\sigma_x^{(i)}-\sigma_x^{(j)}+\sigma_x^{(i)}\sigma_x^{(j)}).
\end{equation}

Below, we discuss problems on the basis
$\{|+\rangle=\frac{1}{\sqrt{2}}(|0\rangle+|1\rangle),
|-\rangle=\frac{1}{\sqrt{2}}(|0\rangle-|1\rangle)\}$. In order to
generate the cluster states of two Josephson charge qubit, we
prepare Josephson charge qubit $i$ in the state
$|\phi\rangle_i=\frac{1}{\sqrt{2}}(|-\rangle_i+|+\rangle_i)$, and
Josephson charge qubit $j$ in the state
$|\phi\rangle_j=\frac{1}{\sqrt{2}}(|-\rangle_j+|+\rangle_j)$, so the
initial state of the system is
$|\phi\rangle_{ij}=\frac{1}{2}(|-\rangle_i+|+\rangle_i)\bigotimes(|-\rangle_j+|+\rangle_j)$.
We assume $i=1$ and $j=2$, according to Hamiltonian $H$ of Eq.
(\ref{4}), we can obtain the following evolutions:
\begin{subequations}
\label{5}
\begin{equation}
|++\rangle_{12}\rightarrow e^{-i\pi t/4\tau}|++\rangle_{12},
\end{equation}
\begin{equation}
|+-\rangle_{12}\rightarrow e^{-i\pi t/4\tau}|+-\rangle_{12},
\end{equation}
\begin{equation}
|-+\rangle_{12}\rightarrow e^{-i\pi t/4\tau}|-+\rangle_{12},
\end{equation}
\begin{equation}
|--\rangle_{12}\rightarrow e^{i3\pi t/4\tau}|--\rangle_{12}.
\end{equation}
\end{subequations}
If we choose $t=\tau$, which can be achieved by choosing switching
time, and perform a single-qubit operation $U=e^{i\pi /4}$, we can
obtain
\begin{subequations}
\label{6}
\begin{eqnarray}|++\rangle_{12}\rightarrow |++\rangle_{12},
\end{eqnarray}
\begin{equation}
|+-\rangle_{12}\rightarrow |+-\rangle_{12},
\end{equation}
\begin{equation}
|-+\rangle_{12}\rightarrow |-+\rangle_{12},
\end{equation}
\begin{equation}
|--\rangle_{12}\rightarrow -|--\rangle_{12}.
\end{equation}
\end{subequations}
 The Eq. (\ref{6}) have actually realized the operation of a
 controlled phase gate. These lead the state of Josephson junction charge qubits $1$ and $2$ to
\begin{eqnarray}
\label{7}
|\phi\rangle_{12}&=&\frac{1}{2}[|-\rangle_1(-|-\rangle_2+|+\rangle_2)+|+\rangle_1(|-\rangle_2+|+\rangle_2)]\nonumber\\
&=&\frac{1}{2}(|-\rangle_1\sigma_z^2+|+\rangle_1)(|-\rangle_2+|+\rangle_2),
\end{eqnarray}
where $\sigma_z^2=|+\rangle_2\langle+|-|-\rangle_2\langle-|$, and
Eq. (\ref{7}) is a standard cluster states of two-qubit. We
generalize the above scheme for generating the cluster states of
two-qubit to the multi-qubit case. We first prepare $N$ ($N\geq2$)
Josephson junction charge qubits in the states
\begin{subequations}
\label{8}
\begin{equation}
|\phi\rangle_1=\frac{1}{\sqrt{2}}(|-\rangle_1+|+\rangle_1),
\end{equation}
\begin{equation}
|\phi\rangle_j=\frac{1}{\sqrt{2}}(|-\rangle_j+|+\rangle_j),
\end{equation}
\end{subequations}
where $j=2,3,\cdot\cdot\cdot,N$. So the total state of $N$ Josephson
charge qubits is
\begin{equation}
\label{9}
|\phi\rangle_{1j}=\frac{1}{2^{N/2}}(|-\rangle_1+|+\rangle_1)\bigotimes_{j=2}^N(|-\rangle_j+|+\rangle_j).
\end{equation}
By choosing the suitable parameters (e.g. $\varepsilon_k(V_Xk),
\overline{E}_{Jk}$, \emph{et al.}), the interaction only occurs
between Josephson charge qubit $i$ and Josephson charge qubit $j$,
while other qubits' interaction don't involved.

Firstly, let Josephson charge qubit 1 only act with Josephson charge
qubit 2 without other qubits' interactions, and make both qubits
undergo the same evolutions as Eq.(\ref{6}). This leads Eq.(\ref{9})
to
\begin{eqnarray}
\label{10}
|\phi\rangle_{1j}&=&\frac{1}{2^{N/2}}(|-\rangle_1\sigma_z^2+|+\rangle_1)\nonumber\\
&(&|-\rangle_2+|+\rangle_2)\bigotimes_{j=3}^N(|-\rangle_j+|+\rangle_j).
\end{eqnarray}
Next, let Josephson charge qubit 2 only act with Josephson charge
qubit 3  without other qubits' interactions. After the same
interaction as Josephson charge qubit 2 with Josephson charge qubit
1, Eq.(\ref{10}) becomes
\begin{eqnarray}
\label{11}
|\phi\rangle_{1j}&=&\frac{1}{2^{N/2}}(|-\rangle_1\sigma_z^2+|+\rangle_1)(|-\rangle_2\sigma_z^3+|+\rangle_2)\nonumber\\
&(&|-\rangle_3+|+\rangle_3)\bigotimes_{j=4}^N(|-\rangle_j+|+\rangle_j).
\end{eqnarray}
From the form of the above states, we can conclude if we let two
Josephson charge qubits interact without other qubits' interactions
every time, step by step, we can obtain the cluster states of
Josephson charge multi-qubit easily. In other words, let Josephson
charge qubit $(j-1)$ only act with Josephson charge qubit $j$
without other qubits' interactions. Thus the cluster states of
Josephson charge $N$ qubits can be obtained
\begin{equation}
\label{13}
|\phi\rangle_{N}=\frac{1}{2^{N/2}}\bigotimes_{j=1}^N(|-\rangle_j\sigma_z^{j+1}|+\rangle_j),
\end{equation}
where $\sigma_z^{N+1}\equiv1$.

Below, we briefly discuss the experimental feasibility of the
current scheme. For the used charge qubit in our scheme, the typical
experimental switching time $\tau^{(1)}$ during a single-bit
operation is about $0.1ns$ \cite{You2002}. The inductance $L$ in our
used proposal is about $30nH$, which is experimentally accessible.
In the earlier design \cite{21-2}, the inductance $L$ is about
$3.6\mu H$, which is difficult to make at nanometer scales. Another
improved design \cite{Makhlin} greatly reduces the inductance $L$ to
$\sim120nH$, which is about 4 times larger than the one used in our
scheme. The fluctuations of voltage source and fluxes result in
decoherence for all charge qubits. The gate voltage fluctuation
plays the dominant role in producing decoherence. The estimated
dephasing time is $\tau_4\sim10^{-4 }s$ \cite{Makhlin}, which allow
in principle $10^6$ coherent single-bit manipulations. Owing to
using the probe junction, the phase coherence time is only about
$2ns$ \cite{Nakamura2,Nakamura2002}. In this setup, background
charge fluctuations and the probe-junction measurement may be two of
the major factors in producing decoherences \cite{You2002}. The
charge fluctuations are principal only in the low-frequency region
and can be reduced by the echo technique \cite{Nakamura2002} and by
controlling the gate voltage to the degeneracy point, but an
effective technique for suppressing charge fluctuations still needs
to be explored.

In summary, we have investigated a simple scheme for generating the
cluster states based on the Josephson charge qubit. We first
introduce the two-qubit case, and then generalize it to multi-qubit
case. Our scheme is simple and easily manipulated, because any two
charge qubits can be selectively and effectively coupled by a common
inductance. The architecture of our proposal is made by present
scalable microfabrication technique. More manipulations can be
realized before decoherence sets in. All the devices in the scheme
are well within the current technology. It is the efficient scheme
for the generation of the cluster states based on the Josephson
charge qubit.

\begin{acknowledgments}
This work is supported by the Key Program of the Education
Department of Anhui Province under Grant No: 2006KJ070A, the Program
of the Education Department of Anhui Province under Grant No:
2006KJ056B and the Talent Foundation of Anhui University.

\end{acknowledgments}

\end{document}